\documentclass[12pt]{article}

\usepackage{amsmath,amsthm,amsfonts,amssymb,amscd}
\def\qed{{\unskip\nobreak\hfil\penalty50
\hskip2em\hbox{}\nobreak\hfil$\square$
\parfillskip=0pt \finalhyphendemerits=0\par}\medskip}

\def\ind{{\mathrm{Ind}}}

\def\Diff{{\mathrm {Diff}}}
\def\Mob{{\rm\textsf{M\"ob}}}

\newtheorem{theorem}{Theorem}[section]
\newtheorem{lemma}[theorem]{Lemma}
\newtheorem{conjecture}[theorem]{Conjecture}
\newtheorem{corollary}[theorem]{Corollary}
\newtheorem{definition}[theorem]{Definition}

\newtheorem{proposition}[theorem]{Proposition}
\newtheorem{remark}[theorem]{Remark}

\def\Diff{{\mathrm {Diff}}}

\def\Mob{{\rm\textsf{M\"ob}}}

\def\res{\!\restriction\!}

\def\A{{\cal A}}

\def\B{{\cal B}}
\def\C{{\cal C}}
\def\D{{\cal D}}

\def\I{{\cal I}}

\def\H{{\cal H}}
\renewcommand{\qed}{\ \hfill $\blacksquare$}

\newcommand{\bdefin}{\begin{definition}}
\newcommand{\edefin}{\end{definition}}

\newcommand{\bdef}{\begin{definition}}
\newcommand{\blem}{\begin{lemma}}
\newcommand{\bprop}{\begin{proposition}}
\newcommand{\bthm}{\begin{theorem}}
\newcommand{\bcor}{\begin{corollary}}
\newcommand{\bconj}{\begin{conjecture}}
\newcommand{\ben}{\begin{equation}}
\newcommand{\een}{\end{equation}}

\newcommand{\ede}{\end{definition}}
\newcommand{\elem}{\end{lemma}}
\newcommand{\eprop}{\end{proposition}}
\newcommand{\ethm}{\end{theorem}}
\newcommand{\ecor}{\end{corollary}}
\newcommand{\econj}{\end{conjecture}}
\newcommand{\brem}{\begin{remark}}
\newcommand{\erem}{\end{remark}}

\newcommand{\ba}{\begin{array}}
\newcommand{\ea}{\end{array}}
\newcommand{\bea}{\begin{eqnarray}}

\title{\huge On singular limits of relative entropies}
\author{
{\sc Feng Xu}\\
Department of Mathematics\\
University of California at Riverside\\
Riverside, CA 92521\\
E-mail: {\tt xufeng@math.ucr.edu}}

\begin{document}
\date{}
\maketitle

\begin{abstract}

In this paper we generalize a key result relating singular limits of certain relative entropies with index  in the setting of conformal nets, which has played an important role recently in the mathematical theory of relative entropies in the context of Conformal Field Theory.

\end{abstract}

\newpage

\section{Introduction}
In the last few years there has been an enormous amount of work by
physicists concerning entanglement entropies in QFT, motivated by
the connections with condensed matter physics, black holes, etc.;
see the references in \cite{Hollands} and  \cite{Wit}  for a partial list of
references, and   \cite{LW},  \cite{LXc},
\cite{Xul} , and \cite{Xul}  for a partial
list of recent mathematical work.

To explain the motivation of this paper, we shall refer the reader to the preliminary section \ref{prelim} for definitions and notations.  Let $\A$ be a conformal net, and $I_1, I_2$  two  intervals with disjoint closures. Let $\omega$ be the state corresponding the vacuum and let $\omega_1\otimes \omega_2$ be the corresponding product  state defined on $\A(I_1)\vee \A(I_2).$ The relative entropy $S_\A(\omega,\omega_1\otimes \omega_2)$ contains rich information about the $\A$, especially when we consider the singular limit where the end points of $I_1$ and $I_2$ are close.  For an example see  Th. 4.2 of \cite{Xul}  where one can read of the central charge and global index of $\A$ from such limits under certain conditions. A key ingredient in the proof of this result is the following. Suppose $\B\subset \A$ is a subnet, then $S_\A(\omega,\omega_1\otimes \omega_2)= S_\B(\omega,\omega_1\otimes \omega_2) + S_\A(\omega, \omega\cdot E_1)$, where $E_1$ is a conditional expectation defined in Section \ref{sin}. The result proved in Th. 4.4 of \cite{Xul} (also stated as Th. 2.20 in \cite{Xutra}) is that when end points of $I_1$ and $I_2$ are close, $S_\A(\omega, \omega\cdot E_1)$ increase to  limit $\ln [\A:\B]$. This is an interesting result connecting relative entropy with index, but under conditions that $\B$ is strongly additive and the index  $[\A:\B]< \infty$. Though these conditions are satisfied for the cases considered in \cite{Xul}, \cite{Xutra}, it is a natural  question to see if one can remove them. In fact in \cite{Casi} the authors argue their computations agree with our Th. 4.4, but in their setting the index $[\A:\B]=\infty$ and 
Th. 4.4  is only proved under the assumption $[\A:\B]< \infty$.  One of the goals of this paper is to remove the assumption  $[\A:\B]< \infty$. In fact we are able to prove such a result in its full generality in Theorem \ref{thm1}.\par
It maybe surprising that one is able to prove such a general result, and we will explain briefly the ideas behind the proof, which is carried out in Section \ref{sin}, and also indicate the preparations for the proof. The proof is divided into three parts. First we consider when $\B\subset \A$ is irreducible, and in this we consider separately the case when  $[\A:\B]< \infty$ and  $[\A:\B]= \infty$. For  $[\A:\B]< \infty$ case, we improve upon the proof of Th. 4.4 of \cite{Xul} by removing strong additivity condition on $\B$. This makes use of 
 M\"obius   covariance and a more streamlined Theorem 2.38 in \cite{Xutra}, which is also stated in Th. \ref{228} for readers' convenience. For  $[\A:\B]=\infty$ case, we use  a result (cf. Lemma \ref{4}) first appeared  \cite{PP} in the setting of type $\mathrm{II}_1$ for subfactors of infinite index, and estimation on relative entropy in Prop. \ref{prop1}. The last case  is when  $\B\subset \A$ is  not irreducible, and we reduce this case to essentially when $\B$ is trivial, and we prove a general result Th. \ref{thm2} about any split  M\"obius  net in Section {\ref{las}. Even though Th. \ref{thm2} is motivated as an intermediate step in the proof of Th. \ref{thm1}, it can be of independent interest. 

\section{Preliminaries}\label{prelim}

\subsection{Spatial derivatives, relative entropy and index theory for general subfactors}\label{sub}

Let  $\psi$ be a  normal state on a von Neumann algebra $M$ acting
on a Hilbert space $H$ and $\phi'$ be a  normal faithful state on
the von Neumann algebra $M'$. The Connes spatial derivative, usually
denoted by $\frac{d\psi}{d\phi'}$,  is a positive  operator (cf.
\cite{Con}) . We will use the simplified notation of \cite{OP} and
write $\frac{d\psi}{d\phi'}= \Delta(\frac{ \psi}{\phi'}).$ If $\psi$
is faithful , we have
$$
\Delta(\frac {\psi}{\phi'})^{it} m\Delta(\frac {\psi}{ \phi'})^{-it}
= \sigma^\psi_t(m), \forall m\in M , \Delta(\frac {\psi}{
\phi'})^{it} m\Delta( \frac{\psi}{ \phi'})^{-it} =
\sigma^{\phi'}_{-t}(m), \forall m\in M'
$$
where $\sigma^\psi_t, \sigma^{\phi'}_{-t}$ are modular
automorphisms.

$$
[D\psi_1: \psi_2]_t:=\Delta(\frac {\psi_1}{ \phi'})^{it}
\Delta(\frac {\psi_2}{ \phi'})^{-it}
$$ is independent of the choice of $\phi'$ and is called {\bf Connes
cocycle}.

Suppose $M$ acts on a Hilbert space $H$ and $\omega$ is a vector
state given by $\Omega\in H$. The relative entropy (cf. 5.1 of
\cite{OP} or \cite{Araki}) in this case is $S(\omega, \phi) = -\langle \ln \Delta
(\phi/\omega') \Omega, \Omega\rangle$ where $\omega' $ is the vector
state on $M'$ defined by vector $\Omega$ and $\Delta
(\phi/\omega'):= \frac{d\phi}{d\omega'}$ is Connes spatial
derivative. When $\Omega$ is not in the support of $\phi$ we set
$S(\omega, \phi)= \infty$.


A list of properties of relative entropies that will be used later
can be found in \cite{OP} (cf. Th. 5.3, Th. 5.15 and Cor. 5.12
\cite{OP}):

\bthm\label{515} (1) Let $M$ be a von Neumann algebra and $M_1$ a
von Neumann subalgebra of M. Assume that there exists a faithful
normal conditional expectation $E$ of $M $onto $M_1$. If $\psi$ and
$\omega$  are states of $M_1$ and $M$, respectively, then $S(\omega,
\psi\cdot E) = S(\omega\res M_1, \psi) + S(\omega, \omega\cdot E);$
\par

(2) Let be $M_i$ an increasing net of von Neumann subalgebras of $
M$ with the property $ (\bigcup_i M_i)''=M$. Then $S(\omega_1\res
M_i, \omega_2\res M_i)$ converges to $ S(\omega_1,\omega_2)$ where
$\omega_1, \omega_2$ are two normal states on $M$; \par

(3) Let $\omega$ and  $\omega_1$ be two normal states on a von
Neumann algebra $M$. If $\omega_1\geq \mu\omega,  $ then $S(\omega,
\omega_1) \leq \ln \mu^{-1}$;

(4) Let $\omega$ and  $\phi$ be two normal  states on a von Neumann
algebra $M$, and denote by  $\omega_1$ and  $\phi_1$ the
restrictions of   $\omega$ and  $\phi$ to  a von Neumann subalgebra
$M_1\subset M$ respectively. Then $S(\omega_1, \phi_1)\leq S(\omega,
\phi)$; \par

(5) Let $\phi$ be a normal faithful state on $M_1\otimes M_2.$
Denote by $\phi_i$ the restriction of $\phi$ to $M_i, i=1,2$. Let
$\psi_i$ be  normal faithful states on $M_i, i=1,2$. Then
$$S(\phi, \psi_1\otimes \psi_2)= S(\phi_1,\psi_1) + S(\phi_2,\psi_2)
+ S(\phi,\phi_1\otimes \phi_2)
$$
\ethm

Let $E:M\rightarrow N$ be a normal faithful conditional expectation
onto a subalgebra $N$. $E^{-1}: N'\rightarrow M'$ is in general an
operator valued weight which verifies the following equation: for
any pair of normal faithful weights $\psi$ on $N$ and $\phi'$ on
$M'$ we have
$$
\Delta(\frac {\psi E}{\phi'}) =\Delta(\frac {\psi }{\phi'E^{-1}})
$$
Kosaki (cf. \cite{Kos}) defined index of $E$, denoted by $\ind E$ to
be  $E^{-1}(1)$.  When $1$ is in the domain of $E^{-1}$, we say that
$E$ has finite index. When both $N, M$ are factors and $E$ has
finite index, we have the (cf. \cite{PP}) Pimsner-Popa inequality
$$E(m)\geq \lambda m, \forall m\in M_+,$$ where
$\lambda= (\ind E)^{-1}$ is the best constant such that the inequality holds. The action of the modular group
$\sigma^{\psi E}_t$ on $N'\cap M$ is independent of the choice of
$\psi$. When $E$ is the minimal conditional expectation such action
is trivial on $N'\cap M$. Also the compositions of minimal
conditional expectations are minimal (cf. \cite{LKo}).

\blem\label{2} Let $\lambda>0$. Then 
$E(m)\geq \lambda m, \forall m\in M_+$, where $M_+$ denote the set of positive elements of $M$, if and only if $||E(m)||\geq \lambda ||m||, \forall m\in M_+$.
\elem
\proof
The proof is given in \cite{BCH} and we include it here for reader's convenience. One direction is trivial. Let us assume that $||E(m)||\geq \lambda ||m||, \forall m\in M_+.$ For any $y\in M$, let $b_n=(E(y^*y)+\frac{1}{n})^{\frac{-1}{2}}, n\geq 1. $ Then we have
$$
b_n y^*y b_n \leq ||b_ny^*yb_n||\leq ||y^*y|| ||b_n^2||\leq \lambda^{-1}  ||E(y^*y)|| ||(E(y^*y)+\frac{1}{n})^{-1}||\leq \lambda^{-1},
$$ and it follows that
$$y^*y\leq b_n^{-1}  \lambda^{-1} b_n^{-1} = (E(y^*y)+\frac{1}{n})  \lambda^{-1} , \forall n\geq 1.$$
Now let $n\rightarrow \infty$ we have finished the proof. \qed

The following result was first proved in the case of type $\mathrm{II}_1$ setting by \cite{PP}. Unfortunately the proof there does not work in the case of type $\mathrm{III}_1$  we are interested in. We give a proof following \cite{BCH}.  See also Page 7 of \cite{Poc} for much more general results. 
\blem\label{4} (1) Suppose that $||E(P)||\geq \lambda >0$ for all nonzero projections $P$ in $M.$  Then $\ind E \leq \frac{1}{\lambda};$\par
(2) If  $\ind E=\infty$ then for any $\epsilon>0$, one can find a non-zero projection $P\in M$ such that $||E(P)||\leq \epsilon$. 
\elem
\proof
Ad (1): By Lemma \ref{2} It is sufficient to check that $||E(m)||\geq \lambda ||m||, \forall m\in M_+$ or equivalently $||E(m)||\geq \lambda , \forall m\in M_+ $  with $||m||=1$. For any $1>\epsilon>0$, let $P$ be the spectral projection of $m$ corresponding to interval $[1-\epsilon,1]$. Since $||m||=1$, $P$ is nonzero, and $m\geq (1-\epsilon) P.$ We have  $E(m)\geq (1-\epsilon)E(P)$, and $||E(m)||\geq (1-\epsilon)||E(P)||\geq  (1-\epsilon) \lambda$ for all 
$1>\epsilon>0.$  Ad(2) : It follows immediately from (1)\qed

\subsection{A result on relative entropy}\label{1.2}

\begin{proposition}\label{prop1}
Suppose that $a\in M_+, ||a||\leq 1,$ and $\omega(a)\geq \epsilon_0, \phi(a)\leq \frac{\epsilon_0^2}{4m}$ where $1> \epsilon_0>0$ , and $m$ is a positive integer. 
Then $S(\omega,\phi)\geq \frac{\epsilon_0\ln m}{2} -e^{-1} + \ln (1-\frac{\epsilon_0}{2m}).$
\end{proposition}

\proof
By assumption the spectrum  $\sigma(a)$ is contained in $[0,1]$. When restricting $\omega, \phi$ to the von Neumann algebra generated by $a$, we can think of $\omega, \phi$  as defined by two probability measures $\mu_1$ and $\mu_2$ on $[0,1]$ such that $\omega(a)=\int_0^1 x du_1(x), \phi(a)=\int_0^1 x du_1(x)$. Let $p$ be the spectral projection of $a$ corresponding to interval $[ \epsilon_0/2,1]$. Then
$$\epsilon_0/2+ \omega(p)\geq \int_0^{\epsilon_0/2} x du_1(x) +  \int_{\epsilon_0/2}^1  x du_1(x)=\omega(a)\geq  \epsilon_0,$$
hence $\omega(p)\geq  \epsilon_0/2$. From 
$$
\frac{\epsilon_0^2}{4m}\geq  \phi(a)= \int_0^1 x du_2(x) \geq \int_{\epsilon_0/2}^1 x du_2(x)\geq \epsilon_0/2 \int_{\epsilon_0/2}^1  du_2(x)= \epsilon_0/2 \phi(p),
$$ we have $\phi(p)\leq \frac{\epsilon_0}{2m}$. 
On the other hand, when restricted to the two dimensional algebra generated by $p$, we have 
$$S(\omega, \phi)= \omega(p)\ln \frac{\omega(p)}{\phi(p)} + \omega(1-p)\ln \frac{\omega(1-p)}{\phi(1-p)}. $$
It follows that 
$$
\omega(p)\ln \frac{\omega(p)}{\phi(p)} + \omega(1-p)\ln \frac{\omega(1-p)}{\phi(1-p)} \geq \frac{\epsilon_0\ln m}{2} -e^{-1} + \ln (1-\frac{\epsilon_0}{2m})$$
where $-e^{-1}$ is the minimal value of the function $x\ln x$ on $(0,1]$. \qed

\subsection{Nets and subnets}\label{net}

\subsubsection{Nets}

This section is contained in \cite{Carp} and \cite{KLM}. We refer to
\cite{Carp} and \cite{KLM} for more details and proofs.

We shall denote by $\Mob$ the M\"obius group, which is isomorphic to
$SL(2,\mathbb R)/\mathbb Z_2$ and acts naturally and faithfully on
the circle $S^1$.

By an interval of $S^1$ we mean, as usual, a non-empty, non-dense,
open, connected subset of $S^1$ and we denote by $\I$ the set of all
intervals.  If $I$ is an interval on the circle on a complex plane
with two end points $a,b$, $r_I:= |b-a|$ is called the length of
$I$.  If $I\in\I$, then also $I'\in\I$ where $I'$ is the interior of
the complement of $I$. Two intervals are {\it disjoint} if their
closure are disjoint. A finite set of intervals are disjoint if any
two different intervals from the set are disjoint.

A {\it  net $\A$ of von Neumann algebras on $S^1$} is a map
\[
I\in\I\mapsto\A(I)
\]
from the set of intervals to the set of von Neumann algebras on a
(fixed) Hilbert space $\H$ which verifies the \emph{isotony
property}:
\[
I_1\subset I_2\Rightarrow  \A(I_1)\subset\A(I_2)
\]
where $I_1 , I_2\in\I$.

A \emph{M\"obius covariant net} $\A$ of von Neumann algebras on
$S^1$ is a net of von Neumann algebras on $S^1$ such that the
following properties $1-4$ hold:
\begin{description}
\item[\textnormal{\textsc{1. M\"obius covariance}}:] {\it There is a
strongly continuous unitary representation $U$ of} $\Mob$ {\it on
$\H$ such that}
\[
U(g)\A(I)U(g)^*=\A({g}I)\ , \qquad g\in \Mob,\ I\in\I \ .
\]
\end{description}
We will write $\alpha_g(a):= U(g)^*a U(g), \forall a\in \A(I).$ 

\begin{description}
\item[$\textnormal{\textsc{2. Positivity of the energy}}:$]
{\it The generator of the rotation one-para\-meter subgroup
$\theta\mapsto U({\rm rot}(\theta))$ (conformal Hamiltonian) is
positive}, na\-me\-ly $U$ is a positive energy representation.

\end{description}
\begin{description}
\item[\textnormal{\textsc{3. Existence and uniqueness of the vacuum}}:]
{\it There exists a unit $U$-invariant vector $\Omega$ (vacuum
vector), unique up to a phase, and $\Omega$ is cyclic for the
von~Neumann algebra $\vee_{I\in\I}\A(I)$}.
\end{description}
%

%
\begin{description}
\item[\textnormal{\textsc{4. locality}}:]
{\it If
$I_1$ and $I_2$ are disjoint intervals,}
\[
[x,y] = 0,\quad x\in\A(I_1), y\in\A(I_2) \ .
\]
\end{description}

A \emph{M\"obius covariant net} $\A$ is nontrivial if $\A(I)\neq \mathbb{C}, \forall I\in \I$. For simplicity we will also call a \emph{M\"obius covariant net} $\A$ simply as a \emph{M\"obius net}.


%

 Now we  recall some definitions from \cite{KLM} . Recall that
${\I}$ denotes the set of intervals of $S^1$. Let $I_1, I_2\in
{\I}$. We say that $I_1, I_2$ are disjoint if $\bar I_1\cap \bar
I_2=\emptyset$, where $\bar I$ is the closure of $I$ in $S^1$. 
Recall that a net ${\A}$ is {\it split} if ${\A}(I_1)\bigvee
{\A}(I_2)$ is naturally isomorphic to the tensor product of von
Neumann algebras ${\A}(I_1)\otimes {\A}(I_2)$ for any disjoint
intervals $I_1, I_2\in {\I}$. ${\A}$ is {\it strongly additive} if
${\A}(I_1)\bigvee {\A}(I_2)= {\A}(I)$ where $I_1\cup I_2$ is
obtained by removing an interior point from $I$. 

A \emph{Conformal  net} $\A$ of von Neumann algebras on $S^1$ is a
net of von Neumann algebras on $S^1$ such that the above  properties
$2-4$ hold, and $1$ is replaces by conformal covariance:
\begin{itemize}
	\item[{}] {\it Conformal covariance}. There exists a projective
	unitary representation $U$ of $\Diff(S^1)$ on $\H$ extending the
	unitary representation of $\Mob$ such that for all $I\in\I$ we have
	\begin{gather*}
		U(g)\A(I) U(g)^*\ =\ \A(gI),\quad  g\in\Diff(S^1), \\
		U(g)xU(g)^*\ =\ x,\quad x\in\A(I),\ g\in\Diff(I'),
	\end{gather*}
\end{itemize}
where $\Diff(S^1)$ denotes the group of smooth, positively oriented
diffeomorphism of $S^1$ and $\Diff(I)$ the subgroup of
diffeomorphisms $g$ such that $g(z)=z$ for all $z\in I'$. By \cite{Wei} a conformal net is split. 


Let $\A$ be a M\"obius  net and $I$ an interval. By removing a point $p$ from $I$, we get  two intervals $I_1, I_2$ contained in $I$ such that  $I_1\cup I_2 \cup \{p \}=I$. Let $\phi$ be a linear functional defined on the algebra $M:=\{a_1a_2, \forall a_1\in \A(I_1),   a_2\in \A(I_2)   \}$ by $\phi(a_1a_2)= \phi(a_1)\phi(a_2) ,  \forall a_1\in \A(I_1),   a_2\in \A(I_2).$ Note that $M$ is dense in $\A(I_1)\vee \A(I_2)	$. 
The following Proposition is essentially contained in \cite{Rob}, and we include a proof in our setting:
\begin{proposition}\label{3}
	Assume that $\A$ is nontrivial. Then $\phi$ can not be extended to a normal state on $\A(I_1)\vee \A(I_2)$.
\end{proposition}
\proof
We will argue by contradiction, assuming that $\phi$ can  be extended to a normal state on $\A(I_1)\vee \A(I_2)$.
By M\"obius covariance we can assume that  $I_1=(-\infty,0), I_2=(0,1)$ and denote by $g_t$ the one parameter family subgroup of $\Mob$ such that $g_t.x= e^{-t}x$. Note that when $t\geq 0$, 
$g_t. I_1\subset I_1, g_t. I_2\subset I_2$, and $g_t.(-1,0)= (-e^{-t},0), g_t.(0,,1)= (0,e^{-t})$. Choose any $a_1\in \A((-1,0)), a_2\in \A((0,1))$.  Since closed balls of $\mathrm{B}(\H)$ is compact in weak topology, we can find a subnet   $\alpha_{g_{t_i}}(a_1a_2)$ that converges weakly to $b$, where $t_i\rightarrow \infty$.   $b$ commutes with $\A((-e^{-t_i}, e^{-t_i}))'$  for  $t_i\rightarrow \infty$, by \cite{LG} we conclude that $b\in \mathbb{C}$. Note that $\omega(\alpha_{g_{t_i}}(a_1a_2)))=\omega(a_1a_2)$ since the vacuum is $\Mob$ invariant, $b=\omega(a_1a_2)$. 
Since $\phi$ is normal on $\A(I_1)\vee \A(I_2)$ by our assumption, we have 
$\lim_{i\rightarrow\infty}\phi(\alpha_{g_{t_i}}(a_1a_2)))= \phi(b)= \omega(a_1)\omega(a_2)= \omega(a_1a_2)$. We have
$$
\langle a_2\Omega, a_1^* \Omega\rangle=\langle \omega(a_2)\Omega, a_1^* \Omega\rangle, \forall a_1  \in \A((-1,0))
$$
Since $\Omega$ is cyclic for $\A((-1,0))$ (cf. \cite{LG})  we conclude that $a_2= \omega(a_2),  \forall a_2 \in \A((0,1))$ which is absurd. \qed

\subsection{Subnets}\label{sub}

Let $\A$ be a  M\"{o}bius net. By a
{\it  M\"{o}bius subnet} (cf. \cite{L1}) we shall mean a map
\[
I\in\I\to\B(I)\subset \A(I)
\]
that associates to each interval $I\in \I$ a von Neumann subalgebra
$\B(I)$ of $\A(I)$, which is isotonic
\[
\B(I_1)\subset \B(I_2), I_1\subset I_2,
\]
and   M\"{o}bius covariant with respect to the representation $U$,
namely
\[
U(g) \B(I) U(g)^*= \B(gI)
\] for all $g\in \Mob$ and $I\in \I.$  When no confusion arises we will  call a {\it  M\"{o}bius subnet} simply a {\it subnet}. Note that by Lemma 13 of \cite{L1} for each $I\in
\I$ there exists a conditional expectation $E_I: \A(I)\rightarrow
\B(I)$ such that $E_I$ preserves the vector state given by the
vacuum of $\A$, and if $E_J$ restricts to $E_I$ if $I\subset J$. 
Let $P$ be the projection onto the closed subspace spanned by $\B(I)\Omega.$ The $E_I(a)P = PaP, \forall a\in \A(I)$. 
\begin{definition}\label{indexd}
Let $\A$ be a  M\"{o}bius  net and  $\B\subset \A $
a subnet. We say   $\B\subset \A $  is of finite index if
$\B(I)\subset \A(I)$is of finite index  for some (and hence all)
interval $I$. The index will be denoted by $[\A:\B].$
\end{definition}
If $\A$ is a conformal net, then $\mathrm{Vir}_\A$ is defined given by $\mathrm{Vir}_\A(I)=  U(\Diff(I))''$.    $\mathrm{Vir}_\A \subset \A$ is also a  subnet and $\mathrm{Vir}_\A$  is  isomorphic to certain  Virasoro net studied in \cite{Carpi}.  We will refer to $\mathrm{Vir}_\A$ as the {\it Virasoro subnet}  of $\A$. 
For a subnet $\B\subset \A $, the relative commutants are defined as $\C(I):=\B(I)'\cap \A, \forall I.$  $\B\subset \A $ is called {\bf irreducible} if $\C(I):=\B(I)'\cap \A(I) = \mathrm{C}$ for some interval $I$. By (1) of the next lemma this condition implies  $\C(I):=\B(I)'\cap \A(I) = \mathrm{C},\forall I. $  The next Lemma follows immediately from definitions. 
\begin{lemma}\label{6}
(1) $U(g) \C(I)U(g^*)=\C(g.I), \forall g\in \Mob; $ 
(2) Assume that $\B$ is strongly additive, then  we have the  \emph{isotony
	property}:
\[
I_1\subset I_2\Rightarrow  \C(I_1)\subset\C(I_2)
\]
where $I_1 , I_2\in\I.$ Denote by $\H_\C$ the closure of $\vee_{I\in \I}\C(I)\Omega$.
Then the map $I\rightarrow \C(I)$ on $\H_\C$ gives an irreducible M\"obius net on $\H_\C.$ 
\end{lemma}

Without strong additivity of $\B$, it is not clear if the the relative commutants $\C(I)$ verify the isotony condition as in Lemma \ref{6}. This problem was discussed in \cite{Koest} and has an affirmative answer under certain conditions. However we can always define a $\D(I):= \B' \cap \A(I),\forall I\in\I$ where $\B:= \vee_{I\in \I}\B(I)$. Note that $\D(I)\subset \C(I), \forall I\in\I,$ and in fact is the largest subnet of $\A$ which verifies  this condition (cf. \cite{Koest}).  By slightly abusing terminology we shall refer to $\D$ as the {\it coset} of $\B\subset \A$. Such a theory has been studied in \cite{Xuc} in concrete models. 
\begin{definition}\label{A}
We say a subnet $\B\ \subset \A$ verifies 	condition A if either $\B\subset \A$ is irreducible or if the net $\D$ defined above is nontrivial. 
\end{definition}
From definition it is clear that if $\B$ is strongly additive, then  $\B\subset \A$ verifies 	condition A.  As another example, assume that a compact group $G$ acts properly on $\A$ such that the fixed point net $\A^G$ is a subnet (cf. \cite{Xup}). By Proposition 2.1 of \cite{Carpi2} $\A^G\subset \A$ is irreducible, and hence verfies 	condition A.  If $\B$ is $\mathrm{Vir}_\A$, then $\B\subset \A$ is irreducible and verifies condition A. Note that if the central charge of $\A$ is greater than $1$ $\mathrm{Vir}_\A$ is not strongly additive (cf. \cite{Bu}).  The following was communicated to us by Sebastiano Carpi:
\begin{lemma}\label{Ca}
Let $\A$ be a conformal net, and let $\B\ \subset \A$ be a subnet. Then $\B\ \subset \A$ verifies condition A.
\end{lemma}
\proof
It is sufficient to check that  if the net $\D$ defined above is trivial, then $\B\subset \A$ is irreducible. If 
the net $\D$ is trivial, by Prop. 2.4 of \cite{C} (also see a brief discussion around equation (51) of \cite{CKLM} ) $\mathrm{Vir}_\A$ is contained in $\B$,  and by 
 Prop. 3.7 of \cite{Carpi} the proof is complete. 
\qed


\section{Singular limits}\label{sin}

Let $\A$ be a conformal net and $I_1, I_2$ are two disjoint intervals. We set
$$\omega_1\otimes \omega (AB)= \langle  \Omega\otimes \Omega,
A\otimes B\ \Omega\otimes \Omega\rangle,\quad \forall A\in
\A(I_1),\  B\in \A(I_2) \ .
$$
Since $\A$ is split (cf. \cite{Wei}), $\omega_1\otimes \omega _2$ extends to a normal state on $\A(I_1)\vee\A(I_2)$.  The mutual information we will compute is
$S(\omega,\omega_1\otimes \omega_2)$. When we wish to emphasize
the underlying net, we will also write the mutual information as
$S_{\A}(\omega,\omega_1\otimes\omega_2).$ When $\B\subset \A$
is a subnet, we write $S_{\B}(\omega,\omega_1\otimes \omega_2)$
the mutual information for the net $\B$ obtained by restricting
$\omega,\omega_1\otimes \omega_2$ from $\A$ to $\B$.  Note that $\B$ is split. Let $F_1$ be the conditional expectation from $ \A(I_1)\vee \A(I_2)$
to  $\B(I_1)\vee \A(I_2)$ such that $F_1(xy)=  E_I(x)y, \forall x\in
\A(I_1),y\in \A(I_2)$ , and  let $E_1$ be the conditional expectation from $ \A(I_1)\vee \A(I_2)$
to  $\A(I_1)\vee \B(I_2)$ such that $E_1(xy)=  x E_I(y), \forall x\in
\A(I_1),y\in \A(I_2).$  First we observe that $\omega\cdot E_1= \omega\cdot F_1$. In fact  $ \forall x\in
\A(I_1),y\in \A(I_2),$  
$$\omega\cdot F_1 (xy)= \langle  E_I(x)y\Omega, \Omega\rangle= \langle  PE_I(x)Py\Omega, \Omega\rangle= \langle  E_I(x)E_I(y)\Omega, \Omega\rangle$$

where we have used $PE_I(x)P= E_I(x)P, PE_I(y)P= E_I(y)P$, and $P\Omega=\Omega$. Similarly $\omega\cdot E_1(xy)= \omega(E_I(x) E_I(y)) $. We also note that $E_{g.I}(\alpha_g(x))= \alpha_g(E_I(x)))$ for all $x\in \A(I)$.  Here interval $I$ is any interval that contains $I_1\cup I_2$. 

Note that $\omega_1\otimes \omega _2\cdot E_I\otimes E_I = \omega_1\otimes \omega _2$ on $\A(I_1)\vee\A(I_2)$, by Th. \ref{515} we have
\begin{equation}\label{12}
S_{\A}(\omega,\omega_1\otimes \omega_2)= S_{\B}(\omega,\omega_1\otimes  \omega_2) +S_{\A}(\omega,\omega\cdot F_1) = S_{\B}(\omega,\omega_1\otimes \omega_2) +S_{\A}(\omega,\omega\cdot E_1)
\end{equation}

It is usually an interesting problem to study the limiting
properties of relative entropies when intervals get close together.
One can find such studies in \S3 and \S4 of \cite{Xul}.
The following Theorem (cf. Th. 2.38 of \cite{Xutra}) plays an important role: 
\bthm\label{228} Assume that $M_n$  is an increasing sequence of
factors act on a fixed Hilbert space, $N_n\subset M_n$ are
subfactors  and $\omega$ is a vector state associated with a vector
$\Omega$. Suppose that $E_n: M_n\rightarrow N_n,  n\geq 1 $ is a
sequence of conditional expectations such that when restricting to
$M_n$, $E_{n+1}=E_n, n\geq 1$, and $\ind E_n = \lambda^{-1} $ is a
positive real number independent of $n$. If strong operator closure
of $\cup_n N_n$ contains $M_1$, then
$$\lim_{n\rightarrow\infty} S(\omega, \omega E_n)= -\ln \lambda
$$\ethm

The following is a generalization of Th. 4.4 of \cite{Xul} (We note that there is a
missing log in Th. 4.4 of \cite{Xul} ):

\bthm\label{thm1} Assume that $\A$ is a conformal net and $\B\subset \A$ is a subnet. 
 Let $I_1$ and $I_2$ be two intervals
obtained from an interval $I$ by removing an interior point, and let
$J_{n}\subset I_2, n\geq 1$ be an increasing sequence of intervals
such that
$$
\bigcup_n J_{n} =I_2,\quad \bar J_n\cap \bar I_1=\emptyset \ .
$$
Let $E_n$ be the conditional expectation from $ \A(I_1)\vee \A(J_n)$
to  $\A(I_1)\vee \B(J_n)$ such that $E_n(xy)= x E_I(y), \forall x\in
\A(I_1),y\in \A(J_n).$ Then
$$
\lim_{n\rightarrow \infty} S(\omega, \omega\cdot E_n)= \ln [\A:\B]\
.
$$
\ethm
\proof
Let $F_n$ be the conditional expectation from $ \A(I_1)\vee \A(J_n)$
to  $\B(I_1)\vee \A(J_n)$ such that $F_n(xy)=  E_I(x)y, \forall x\in
\A(I_1),y\in \A(J_n).$   

By equation (\ref{12})  $S(\omega, \omega\cdot E_n)= S(\omega, \omega\cdot F_n)$ and we find it more convenient to work with $S(\omega, \omega\cdot F_n)$.  We also note that $E_{g.I}(\alpha_g(x))= \alpha_g(E_I(x)))$ for all $x\in \A(I)$.  
 Let $\mu= \lim_{n\rightarrow \infty} S(\omega, \omega\cdot F_n). $ By Mobius covariance it is clear that $\mu$ is independent of the choice of $I_1, I_2$. We can assume that $I_1=(-\infty,0), I_2=(0,1)$ and we can choose $J_n=(1/n^2,1)$. Applying dilation  which is multiplication  by positive integer $n$ to $I_1, J_n$ we can now assume  $I_1=(-\infty,0), J_n=(1/n,n)$.  We will compute $\mu$ using  $I_1=(-\infty,0), J_n=(1/n,n)$. \par
Case I: $\B(I_1)'\cap \A(I_1)= \mathbb{C}$, and  $[\A:\B]<\infty$. We will apply Th. \ref{228} in this case for $M_n =\A(I_1)\vee \A(J_n), N_n=\B(I_1)\vee \A(J_n)$. Note that $\vee_n N_n 
=  \B(I_1) \vee_n \A(J_n)= \B(I_1)\vee \A((0,\infty)).$ Since $\A((0,\infty)=\A(I_1)'$ and $\B(I_1)'\cap \A(I_1)= \mathbb{C}$, we conclude that $M_1\subset \vee_n N_n$ and Th. \ref{228}  implies our Theorem in this case. 

Case II: $\B(I_1)'\cap \A(I_1)= \mathbb{C}$, and  $[\A:\B]=\infty$. Let $m, 1>\epsilon_0>0$ be as in Prop.  \ref{prop1}. Since $\ind E_{I_1}=\infty$, by Lemma \ref{4} we can choose a projection $e_1\in \A(I_1)$ such that $||F_1(e_1)||< \frac{\epsilon_0^2}{4m}.$ On the other hand since  $\A(I_1)$ is a type $\mathrm{III}$ factor we can find unitary $u\in \A(I_1)$ such that $\omega(u^*e_1u)> 2\epsilon_0$. 
Since  $M_1\subset \vee_n N_n,$  we can find $u_n$ in the unit ball of $N_n$ such that $u_n\rightarrow u, u_n^*\rightarrow u^*$ in strong operator topology for some sequence of $n$ which goes to infinity.  Here we note that $\H$ is separable due to the split property of $\A$ (cf. \cite{Wei}). Let $P_n= u_n e_1 u_n^*$. Then $P_n\in M_n$ converges to $ue_1u^*$ strongly. It follows that when $n$ is large enough we have 
$$\omega(P_n)> \epsilon_0, \omega F_n (P_n)=  \omega (u_n F_1(e_1)u_n^* )\leq ||F_1(e_1)|| <\frac{\epsilon_0^2}{4m}. $$
By Prop. \ref{prop1} we have $S(\omega, \omega\cdot F_n)\geq \frac{\epsilon_0\ln m}{2} -e^{-1} + \ln (1-\frac{\epsilon_0}{2m})$. Therefore $\mu\geq  \frac{\epsilon_0\ln m}{2} -e^{-1} + \ln (1-\frac{\epsilon_0}{2m})$ for any $m\geq 1$. Let $m$ go to $\infty$ we have shown $\mu=\infty$ in this case.

Case III: Assume  that $\B\subset \A$ is not irreducible. Then the index $[\A,\B]=\infty $  by Lemma 14 of \cite{L1}. By Lemma \ref{Ca} the coset $\D=\B'\cap \A$ is nontrivial.  We note that for any interval $J$, $\B(J)\vee \D(J)$ is isomorphic to $\B(J)\otimes \D(J)$ (cf. \cite{Koest}), and $\omega$ on $\B(J)\vee \D(J)$ is a tensor product of its restriction on $\B(J)$ and $\D(J)$. Restricting $\omega, \omega\cdot F_n$ to $\B\vee \D$, by (5) of Th. \ref{515} we have 
$S_{\B\vee \D}(\omega, \omega\cdot F_n)= S_\D(\omega, \omega_{I_1}\otimes \omega_{J_n}).$ Since $\A$ is split, the subnet $\D$ is  also split.  Our Theorem follows from Theorem \ref{thm2}.\qed

The case considered in \cite{Casi} corresponds to the case when $\A$ is a strongly additive net, and $\B$ is the fixed point net of $\A$ under the action of a compact group.  Th. \ref{thm1} holds for such $\B\subset \A$. 
\begin{remark}
For simplicty we have restricted to local nets instead of graded nets in this paper.  Th. \ref{thm1} can be easily generalized to the $\mathrm{Z}_2$ graded setting, for example by simply passing to the $\mathrm{Z}_2$ fixed point subnet which is local. 
\end{remark}
\subsection{Proof of Th. \ref{thm2}}\label{las}
\bthm\label{thm2} Assume that $\A$ is a M\"obius net with split properrty .  Let $I_1$ and $I_2$ be two intervals
obtained from an interval $I$ by removing an interior point, and let
$J_{n}\subset I_2, n\geq 1$ be an increasing sequence of intervals
such that
$$
\bigcup_n J_{n} =I_2,\quad \bar J_n\cap \bar I_1=\emptyset \ .
$$
 Let $\phi_n$ be the tensor states such that 
$\phi_n(xy)=\omega(x)\omega(y),  \forall x\in
\A(I_1),y\in \A(J_n).$ 

Then
$$
\lim_{n\rightarrow \infty} S_{\A(I_1)\vee \A(J_n)}(\omega, \phi_n)= \infty. \
$$
\ethm
\proof

We will argue by contradiction. Note that by monotonicity $S_{\A(I_1)\vee \A(J_n)}(\omega, \phi_n)$ increase with $n$. For simplicity we will drop the subscript $\A(I_1)\vee \A(J_n)$ in the following. Assume that $$
\lim_{n\rightarrow \infty} S(\omega, \phi_n)= \mu< \infty. \
$$
Under this assumption we will show that $\phi_n$ can be extended to a normal state on $\A(I_1)\vee \A(I_2)$, which  contradicts  Prop. \ref{3}. Denote by $M:=\cup_n \A(I_1)\vee \A(J_n)$. Since $\A(I_1)\vee \A(J_n)$ is an increasing sequence of von Neumann algebras, $M$ is a strongly dense $*$ subalgebra of $\A(I_1)\vee \A(I_2)$. Let $\phi$ be the linear functional on $M$ which  restricts to  $\phi_n$ on $\A(I_1)\vee \A(J_n)$. 
Note that $\phi$ is faithful on $M$ since it is faithful on each $\A(I_1)\vee \A(J_n).$

We apply GNS construction to $(M,\phi)$. Let $H_\phi$ be the corresponding Hilbert space. We will write $\tilde{m}$ for the operator on $H_\phi$ which represents the action of $M$, and $\hat{m}$ the vector corresponding to $m$ in  $H_\phi$. We note that $\hat{M}$ is dense in $H_\phi$ and similarly $M\Omega$ is dense in $\H$. Denote by $W$ the  strong operator closure of $\tilde{M}$ on  $H_\phi$. 
Let $F(\tilde{m})=m, \forall m \in M$.  $F$ is an isomorphism on each $\A(I_1)\vee \A(J_n)$. We need to show that $F$ extends to a homorphism  from $W$ to $\A(I_1)\vee \A(I_2)$. First we show we can extend $F$ to 
$W$. It is sufficient to do this on the unit ball $(W)_1$ of $W$. Since the unit ball of $(\tilde{M})_1$ is dense in  $(W)_1$  by \S2 of \cite{Tak}, it is sufficient to show that if a sequence $\tilde{a}_n \in (\tilde{M})_1$ converges  strongly to $0$, then  $a_n$ converges strongly to $0$ for any $||a_n|\leq 1, \forall n.$ Suppose that $a_n$ does not converge strongly to $0$, then one can find a vector $b\Omega\in \H$  for some $b\in (M)_1,$  a positive number $\epsilon_0$ such that we have a subsequence $a_{n_k}$  such that $\langle a_{n_k} b \Omega, a_{n_k}b\Omega\rangle = \omega(b^* a_{n_k}^*a_{n_k} b )\geq \epsilon_0$. On the other hand since $\tilde{a}_n$ converges  strongly to $0$,  for any positive integer $m$ we can find  $n_k$  large enough such that $\langle a_{n_k} \hat{b}, a_{n_k}\hat{b}\rangle =\phi(b^* a_{n_k}^*a_{n_k} b ) \leq \frac{\epsilon_0^2}{4m}$. By Prop. \ref{prop1}  we have
$S(\omega,\phi_{n_k})\geq \frac{\epsilon_0\ln m}{2} -e^{-1} + \ln (1-\frac{\epsilon_0}{2m}).$ It follows that $\mu \geq S(\omega,\phi_{n_k})\geq \frac{\epsilon_0\ln m}{2} -e^{-1} + \ln (1-\frac{\epsilon_0}{2m})$ for any positive integer $m$, a contradiction. Hence $F$ extends to a normal homorphism from $W$ to $\A(I_1)\vee \A(I_2)$. Since $M$ is dense in $\A(I_1)\vee \A(I_2),$ it follows that $F(W)=A(I_1)\vee\A(I_2)$. Let $W_1$ be the kernel of $F$. This is a two sided weakly closed $*$ ideal of $W$, so there is a central projection $Q\in W$ such that $W_1=QW$. When restricting to $(1-Q)W,$ $F$ is an isomorphism of $(1-Q)W$ and 
$A(I_1)\vee \A(I_2)$ and hence $\phi$ extends  to a normal state on  $\A(I_1)\vee \A(I_2).$\qed


\vspace{0.4cm}
\noindent {\bf Acknowledgements}

The author is grateful to Professor Sebastiano Carpi for discussions about Lemma \ref{Ca}.

\noindent {\bf Data Availability Statement}

All data generated or analyzed during this study are included in this article.

\vspace{0.4cm}

\noindent {\bf Declarations}

\vspace{0.4cm}

\noindent {\bf Conflict of Interest Statement}.

The author declares that there are no conflict of interest regarding the publication of this manuscript.

{\footnotesize

\end{document}